\documentstyle[12pt,epsf]{article}

\newskip\zatskip \zatskip=0pt plus0pt minus0pt
\def\matth{\mathsurround=0pt}
\def\gtrsim{\mathrel{\mathpalette\atversim>}}
\def\lstsim{\mathrel{\mathpalette\atversim<}}
\def\atversim#1#2{\lower0.7ex\vbox{\baselineskip\zatskip\lineskip
  \zatskip
  \lineskiplimit 0pt\ialign{$\matth#1\hfil##\hfil$\crcr#2\crcr\sim
  \crcr}}}

\begin{document}
\topskip 2cm 
\begin{titlepage}

\hspace*{\fill}\parbox[t]{4cm}{MSUHEP-90122
\\ Jan 24, 1999 \\ hep-ph/9901397}

\vspace{2cm}

\begin{center}
{\large\bf Rapidity-Separation Dependence and the Large 
Next-to-Leading Corrections to the BFKL Equation} \\
\vspace{1.5cm}
{\large Carl R. Schmidt} \\
\vspace{.5cm}
{\sl Department of Physics and Astronomy\\
Michigan State University\\
East Lansing, MI 48824, USA}\\
\vspace{1.5cm}
\vfil

\begin{abstract}

Recent concerns about the very large next-to-leading logarithmic (NLL)
corrections to the BFKL equation are addressed by the introduction of
a physical rapidity-separation parameter $\Delta$.
At the leading logarithm (LL) this parameter enforces the constraint 
that successive emitted 
gluons have a minimum separation in rapidity, $y_{i+1}-y_i>\Delta$.  
The most significant effect is to reduce the BFKL Pomeron intercept from 
the standard result as $\Delta$ is increased from 0 (standard BFKL).
At NLL this $\Delta$-dependence is compensated by a modification of 
the BFKL kernel, such that the total dependence on 
$\Delta$ is formally next-to-next-to-leading logarithmic.  
In this formulation, as long as $\Delta\gtrsim2.2$ (for 
$\alpha_{s}=0.15$): (i) the NLL BFKL pomeron intercept is stable with 
respect to variations of $\Delta$, and (ii) the NLL correction is 
small compared to the LL result.  Implications 
for the applicability of the BFKL resummation to phenomenology are 
considered.

\end{abstract}
\end{center}

\end{titlepage}

\section{Introduction}
\label{sec:0}

Recently, the long-awaited next-to-leading-logarithmic (NLL) 
corrections to the Balitsky-Fadin-Kuraev-Lipatov (BFKL) equation 
\cite{FKL}-\cite{bal} have been completed \cite{nll, fadin}%
\footnote{These NLL corrections rely on the intermediate results of 
many individuals \cite{real}-\cite{bds}.  A partially-independent 
confirmation of the final result can be found in \cite{cc}.}.  
The BFKL equation is used to resum 
the large logarithms in Quantum Chromodynamics (QCD) of the type 
$\ln(\hat s/|\hat t|)$, 
where $\hat s$ is the center-of-mass energy-squared of the partonic
scattering process and $\hat t$ is of the order of the momentum transfer 
in the process.  The most obvious result of this NLL 
calculation is the correction to the ``BFKL Pomeron'' intercept $\omega_{P}$, 
which describes the rise of the total cross section with $\hat s$.  The 
asymptotic form of the high-energy partonic cross section predicted by the 
leading-logarithmic (LL) BFKL resummation is
\begin{equation}
	\hat \sigma\approx {1\over\hat t}
	\left({\hat s\over |\hat t|}\right)^{\omega_{P}-1}\ ,
	\label{regtraj}
\end{equation}
where 
\begin{equation}
	\omega_{P}-1\ =\ {4N_{c}\alpha_{s}\ln{2}\over\pi}
	\label{llreg}
\end{equation}
with $N_{c}=3$ the number of colors.  At NLL one obtains 
the ${\cal O}(\alpha_{s}^{2})$ correction to $\omega_{P}-1$.  
Unfortunately, the NLL correction to this parameter is large and negative 
\cite{nll,fadin}.  In addition the saddle-point approximation, which 
was used to 
obtain the asymptotic form of equation (\ref{regtraj}) at LL, gives a 
cross section which is no longer strictly positive-definite at NLL 
\cite{ross}. These and other problems have led some researchers 
to call into question the reliability of the NLL 
BFKL resummation for phenomenological applications \cite{levin,ball}.  
At the very least one would like to know what is the meaning of these 
large NLL corrections.  Can one understand them and can one control them?

On inspection,
the NLL BFKL equation and solution as presented by Fadin and Lipatov
 appear to be free of any arbitrary parameters.
Let us compare this with another logarithmic resummation, that of a 
fixed-order perturbative cross section, using a
running coupling in the $\overline{\rm MS}$ scheme.  At the Born-level 
one calculates the cross section with the running coupling 
evaluated to LL accuracy.  This cross section depends on an arbitrary 
parameter $\mu$, the scale of the running coupling, which determines 
the size of the resummed logarithm and is usually chosen to 
be of the order of some relevant scale in the scattering process.  At 
next-to-leading order (NLO) in the matrix element calculation, one uses the
NLL calculation of the running coupling, and the dependence on $\mu$ 
cancels effectively to one higher order in the perturbation expansion.
Thus, if all is well, the dependence on $\mu$ is reduced in the NLO 
calculation. In fact, the dependence on this parameter is often interpreted 
as an estimate of the theoretical uncertainty due to higher order corrections.

This leads one to ponder whether there might be a similar 
arbitrary-parameter dependence hidden in the BFKL resummation.  
Here the large logarithms that are being resummed arise from the 
integration over the rapidities of the real and virtual gluons in the 
squared amplitude.  To LL accuracy, the exact range of these rapidity 
integrations is not precisely defined.  One could reduce the range of 
integration by a small amount and still be within the validity of 
the LL approximation.  The excluded rapidity range will then resurface
as part of the NLL correction, and the dependence on this separation 
between LL and NLL should vanish, up to contributions which are 
formally next-to-next-to-leading-logarithmic (NNLL).

With this in mind we consider a modification of the BFKL equation, 
where the integration over gluon rapidities $y_{i}$ is subject to the 
constraint that $y_{i+1}-y_{i}>\Delta$, with the parameter $\Delta $ 
assumed to be much less than the total rapidity interval%
\footnote{This idea has been considered before at the LL level in 
Refs.~\cite{bo,liptalk}, where the modification of the LL BFKL pomeron 
intercept was found.}.  
This reduces to the standard BFKL equation for $\Delta=0$. 
There are several reasons why a nonzero value of $\Delta$ might be 
preferable.  First, in the derivation of the LL BFKL equation the 
assumption of multi-Regge kinematics was used in extracting the 
BFKL amplitudes.  That is, contributions which are formally 
${\cal O}(e^{-|y_{i}-y_{j}|})$ are neglected in the amplitudes.  Thus, 
the region $y_{i+1}\sim y_{i}$ is precisely 
the region where this approximation is worst.  It makes sense 
to shift these regions of integration over rapidity into the NLL 
corrections where the assumption of multi-Regge kinematics is relaxed.
Second, in certain processes such as dijet production at hadron 
colliders the BFKL calculations greatly overestimate the cross 
section due to the lack of energy-longitudinal momentum conservation 
at LL \cite{carl,os}.  By keeping the gluons away from the 
ends of the rapidity interval, one can reduce this effect.  Again, at 
NLL these regions of the rapidity integration would be added back in, 
but with energy-longitudinal momentum conservation preserved for the 
first gluon in the ladder.  Finally, we note 
that the large negative NLL corrections suggest that the LL 
prediction should be reduced, which naturally occurs for $\Delta>0$.

In the remainder of this paper we explore the consequences of this 
modification of the BFKL equation.  In section 2 we solve the 
BFKL equation with the constraint on the rapidity separation at LL and 
show how this affects the LL prediction for the BFKL Pomeron intercept.
In section 3 we show how the constraint on the rapidity separation in 
the BFKL equation is translated into a modification of the small-$x$ 
resummation of the gluon-gluon 
splitting functions.  In section 4 we consider the constrained
BFKL equation at NLL.  Using the fact that the exact high-energy cross 
section should have no dependence on the arbitrary parameter $\Delta$,
 we can obtain the 
$\Delta$-dependence of the NLL BFKL kernel.  This result 
is then combined with the standard NLL corrections to obtain the NLL 
prediction for the BFKL Pomeron intercept as a function of the 
parameter $\Delta$.  Finally, in section 5 we discuss the 
phenomenological consequences of our results, and we present our 
conclusions.  

\section{BFKL Equation with Constraint on Rapidity Separations}
\label{sec:1}

In order to be precise, throughout this paper we use the term rapidity to
mean the physical rapidity, defined by
\begin{equation}
	y\ =\ {1\over2}\ln{E+p_{z}\over E-p_{z}}\ .
	\label{rap}
\end{equation}
Thus, in the 
{\sl multi-Regge kinematics}, which presumes that the produced gluons are
strongly ordered in rapidity and have comparable transverse momenta,
\begin{eqnarray}
y_{b} \ll y_{1}\ll y_{2}\ll &\cdots& \ll y_{a}\nonumber\\
|q_{b\perp}|\simeq|q_{1\perp}|\simeq|q_{2\perp}|\simeq&\cdots&
\simeq|q_{a\perp}|\simeq
|k_{1\perp}|\simeq|k_{2\perp}|\simeq\cdots\, ,\label{mrk}
\end{eqnarray}
the rapidity intervals are given by 
\begin{equation}
	y_{ij}\ =\ y_{i}-y_{j}\ \simeq\ \ln{s_{ij}\over|k_{i\perp}|\,|k_{j\perp}|}
	\ ;\qquad\qquad y_{i}\gg y_{j}\ .
	\label{rapint}
\end{equation}
In these equations, the $k_{i\perp}$ are the transverse momenta of the 
emitted gluons, the 
$q_{i\perp}$ are the transverse momenta of the reggeized gluons 
exchanged in the $t$-channel, and $s_{ij}=(k_{i}+k_{j})^{2}$.  A 
typical diagram which is used to build the BFKL ladder at LL is shown 
in figure 1.  

\begin{figure} 
\vskip1.0cm
\epsfysize=3.5cm
\centerline{\epsffile{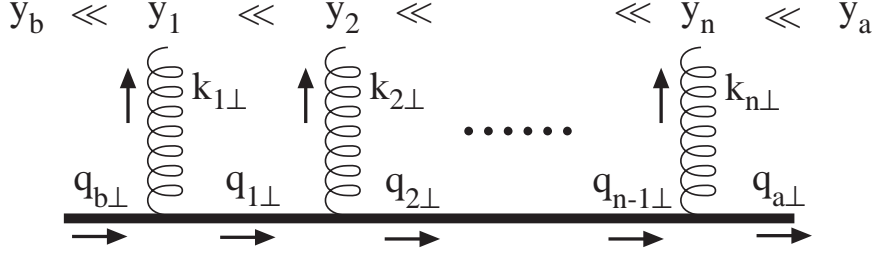}}
\vskip-0.0cm
\vskip6pt
\caption{A BFKL ladder diagram.  The heavy line represents the reggeized 
gluon exchanged in the $t$-channel.}
\label{fig1}
\vskip-1.0cm
\end{figure}

We now consider the modified BFKL equation at LL, given by
\begin{eqnarray}
f( y_{a}- y_{b},q_{a\perp},q_{b\perp}) &=&  
{1\over2}\delta^{2}(q_{a\perp}-q_{b\perp})\Theta( y_{a}- y_{b}-\Delta)
\nonumber\\
&&
+\ 
{\bar\alpha_{s}\over\pi}\int_{ y_{b}+\Delta}^{ y_{a}-\Delta} 
d y\int{d^{2}k_\perp
\over k_{\perp}^{2}}\Biggl[
f( y- y_{b},q_{a\perp}\!+\!k_\perp,
q_{b\perp})\nonumber\\
&&\qquad\qquad\qquad
-{q_{a\perp}^{2}\over k_{\perp}^{2}+(q_{a\perp}+k_{\perp})^{2}}
f( y- y_{b},q_{a\perp},q_{b\perp})\Biggr]\label{bfkl} \ ,
\end{eqnarray}
where $\bar\alpha_{s}=\alpha_{s}N_{c}/\pi$.  The function 
$f( y_{ab},q_{a\perp},q_{b\perp})$ is the BFKL Green's function 
which describes the flow of transverse momentum from $q_{b\perp}$ to 
$q_{a\perp}$ by the emission of real and virtual gluons along the rapidity 
interval $y_{ab}=y_{a}-y_{b}$.  The $\Delta$-dependence in this equation 
just enforces the constraint $y_{i+1}-y_{i}>\Delta$ for each 
successive emitted gluon in the ladder.  To be 
consistent, the constraint is applied to the integrals in rapidity which 
are associated with both the real and virtual gluons.  For $\Delta=0$ 
this equation just reduces to the standard BFKL equation, but eq.~(\ref{bfkl}) 
gives an equally valid LL resummation for any $\Delta\ll y_{ab}$.

We can easily solve eq.~(\ref{bfkl}) in the same manner as the original BFKL 
solution.  First, perform a Mellin transform on this equation, defining
\begin{equation}
	f_{\omega}(q_{a\perp},q_{b\perp})\ =\ 
	\int_{\Delta}^{\infty}dy\,e^{-\omega(y-\Delta)}
	f( y,q_{a\perp},q_{b\perp}) .
	\label{mellin}
\end{equation}
This gives the equation
\begin{eqnarray}
\lefteqn{\omega\,f_{\omega}( q_{a\perp},q_{b\perp}) \ =\ 
{1\over2}\delta^{2}(q_{a\perp}-q_{b\perp})}
\nonumber\\
&&\qquad
+\ e^{-\omega\Delta}\,
{\bar\alpha_{s}\over\pi}\int{d^{2}k_\perp
\over k_{\perp}^{2}}\Biggl[
f_{\omega}( q_{a\perp}\!+\!k_\perp,
q_{b\perp})
-{q_{a\perp}^{2}\over k_{\perp}^{2}+(q_{a\perp}+k_{\perp})^{2}}
f_{\omega}( q_{a\perp},q_{b\perp})\Biggr]\label{bfklmel} \ ,
\end{eqnarray}
The integral operator over the transverse momentum space has the same 
eigenfunctions and eigenvalues as the original BFKL equation, so 
we can immediately write down the solution to this equation:
\begin{equation}
f_{\omega}(q_{a\perp},q_{b\perp})\ =\ {1\over (2\pi)^2 }
\sum_{n=-\infty}^{\infty} e^{in\phi}\, 
\int_{-\infty}^{\infty} d\nu\, 
{(q_{a\perp}^2)^{-1/2+i\nu}\,(q_{b\perp}^{2})^{-1/2-i\nu}\over 
\omega-\omega(n,\nu)\,e^{-\omega\Delta}}\ ,\label{solc}
\end{equation}
where $\phi=\phi_{a}-\phi_{b}$ and
the eigenvalue of the integral operator is
\begin{equation}
	\omega(n,\nu)\ =\ 
	2\bar\alpha_{s}\left[\psi(1)-{\rm Re}\,\psi\left({|n|+1\over2}+i\nu\right)\right]
	\ ,
	\label{eig}
\end{equation}
where $\psi$ is the logarithmic derivative of the gamma function.

By performing the inverse Mellin transform, we obtain the BFKL Green's 
function as a function of the rapidity interval $y$:
\begin{equation}
f(y,q_{a\perp},q_{b\perp})\ =\ {1\over (2\pi)^2 }
\sum_{n=-\infty}^{\infty} e^{in\phi}\, 
\int_{-\infty}^{\infty} d\nu\, 
{(q_{a\perp}^2)^{-1/2+i\nu}\,(q_{b\perp}^{2})^{-1/2-i\nu}
e^{\tilde\omega(n,\nu)(y-\Delta)}\over 
1+\tilde\omega(n,\nu)\Delta}\ ,\label{sold}
\end{equation}
where $\tilde\omega(n,\nu)$ is a solution of the equation 
\cite{bo,liptalk} 
\begin{equation}
	\tilde\omega(n,\nu)\ =\ \omega(n,\nu)e^{-\tilde\omega(n,\nu)\Delta}\ .
	\label{tilo}
\end{equation}

It is an interesting exercise to expand the equation (\ref{sold}) 
order-by-order as a power series in $\bar\alpha_{s}$.  This is done 
using the formula:
\begin{equation}
{e^{\tilde\omega(n,\nu)(y-\Delta)}\over 
1+\tilde\omega(n,\nu)\Delta}\ =\ \sum_{m=0} {\omega(n,\nu)^{m}
[y-(m+1)\Delta]^{m}\over m!}\ .
	\label{expom}
\end{equation}
The factor $[y-(m+1)\Delta]^{m}/m!$ in (\ref{expom}) 
is exactly the phase space in rapidity for $m$ 
intermediate gluons, subject to the constraint $y_{i+1}-y_{i}>\Delta$. 
In fact, for a given rapidity interval $y$ the series should actually 
be truncated at the largest value of $m$ for which $[y-(m+1)\Delta]>0$, 
because one cannot put any more gluons in the rapidity interval and still 
obey the constraint.  However, we also note that the power series  
converges only asymptotically to the analytic expression on the left-hand 
side of (\ref{expom}), and the best approximation is obtained by the 
truncated series.  Thus, as $y/\Delta$ is increased, the analytic 
solution and the truncated series become arbitrarily close.  

For asymptotically-large $y$ we can perform the integration over $\nu$
in eq.~(\ref{sold}) using 
the saddle-point approximation.  The eigenvalue $\tilde\omega(n,\nu)$ is 
largest for $n=0$ and is strongly peaked near $\nu=0$.  Thus, we may
keep only the first term in the Fourier series in $\phi$, and we can 
expand 
\begin{equation}
	\tilde\omega(0,\nu)\ =\ \tilde A-\tilde B\nu^{2}+\ \cdots\ .
	\label{nuexp}
\end{equation}
The coefficients $\tilde A$ and $\tilde B$ are related to the standard 
BFKL saddle-point coefficients,
\begin{equation}
	A\ =\ 4\bar\alpha_{s}\ln{2};\qquad\qquad B\ =\ 
	14\bar\alpha_{s}\zeta(3)\ ,
	\label{ab}
\end{equation}
as solutions to the equations
\begin{eqnarray}
	\tilde A & = & Ae^{-\tilde A\Delta}\nonumber  \\
	\tilde B & = & {B\,(\tilde A/A)\over1+\tilde A\Delta}\ .
	\label{tilab}
\end{eqnarray}
Evaluating the integral over $\nu$ in the saddle-point approximation 
gives
\begin{equation}
	f(y,q_{a\perp},q_{b\perp})\ =\ {e^{\tilde A(y-\Delta)}\over4\pi
	|q_{a\perp}||q_{b\perp}|\,
	(1+\tilde A\Delta)\,\sqrt{\tilde B\pi\tilde y}}\,
	\exp{\left(
	-{\ln^{2}(q_{a\perp}^{2}/q_{b\perp}^{2})\over 4\tilde B\tilde y}
	\right)}\ ,
	\label{asymf}
\end{equation}
where
\begin{equation}
	\tilde y\ =\ y-\Delta{2+\tilde A\Delta\over1+\tilde A\Delta}\ .
	\label{tily}
\end{equation}

Using the relation $y=\ln(\hat s/|q_{a\perp}||q_{b\perp}|)$, we 
recognize that the quantity $\tilde A$ is related to the BFKL Pomeron 
intercept ($\tilde A=\omega_{P}-1$).  For 
$\Delta\ge 0$ it is bounded by $0<\tilde A\le A$.  In table 1 we give the
magnitude of $\tilde A$ for several representative 
values of the rapidity-separation parameter $\Delta$ with 
$\alpha_{s}=0.15$.  For $\Delta=2$, the prediction for $\tilde A$ 
is reduced substantially, from 0.397 to 0.244.  Recall that any of 
these predictions are equally valid in the LL approximation, assuming that 
$y\gg\Delta$.
\begin{table}[t]
	\centering
	\vskip2.0cm
	\begin{tabular}{|c||c|c|c|c|c|}
		\hline
		$\phantom{\Biggl[}\Delta$ & 0 & 0.5 & 1 & 1.5 & 2  \\
		\hline
		$\phantom{\Biggl[}\tilde A$ & 0.397 & 0.336 & 0.296 & 0.266 & 0.244  \\
		\hline
	\end{tabular}
	\caption{BFKL Pomeron intercept $\tilde A=\omega_{P}-1$ at LL for several 
	values of rapidity-separation parameter $\Delta$,
	with $\alpha_{s}=0.15$.}
	\vskip-1.5cm
	\label{pslope}
\end{table}

\section{$\Delta$-dependent Gluon Anomalous Dimension}
\label{sec:2}

We can use eqs.~(\ref{solc}) and (\ref{sold}) to obtain 
the gluon-gluon splitting function at small $x$ by introducing 
a $k_{\perp}$-dependent gluon distribution function ${\cal 
F}(x,k_{\perp})$, which is related to the standard gluon distribution 
function $g(x,M^{2})$ via
\begin{equation}
	xg(x,M^{2})\ =\ \int d^{2}k_{\perp}{\cal 
	F}(x,k_{\perp})\,\Theta(M^{2}-k_{\perp}^{2})\ ,
	\label{ktsplit}
\end{equation}
where $M$ is the factorization scale%
\footnote{For a more rigorous discussion of 
$k_{\perp}$-factorization, see Ref.~\cite{cathaut}.}.
The function ${\cal 
F}(x,k_{\perp})$ satisfies the inhomogeneous BFKL 
equation, so that it has a general solution of the form
\begin{equation}
	{\cal F}(x,k_{\perp})\ =\ \int_{x}^{1}{dz\over 
	z}f(\ln{z/x},q_{\perp},k_{\perp})\, {\cal 
	F}_{0}(z,q_{\perp})\bigg|_{q_{\perp}\rightarrow0}\ ,
	\label{bare}
\end{equation}
where ${\cal F}_{0}(z,q_{\perp})$ can be considered a ``bare'' gluon 
distribution 
and $q_{\perp}$ is used as an infrared cutoff.  The splitting 
function can then be obtained by taking the derivative $\partial 
g(x,M^{2})/\partial\ln{M^{2}}$.  In practice, it is more convenient to 
work with the moments of these equations to obtain the gluon-operator
anomalous dimension, which is related to the gluon-gluon splitting 
function by
\begin{equation}
	\gamma_{N}\ =\ {\alpha_{s}\over2\pi}\int_{0}^{1}dx 
	\,x^{N-1}P_{gg}(x)\ .
	\label{andim}
\end{equation}
Following this line of argument, we obtain the anomalous dimension as 
the implicit solution to the equation \cite{bo}
\begin{equation}
	{\bar\alpha_{s}e^{-(N-1)\Delta}\over N-1}\left[2\psi(1)-\psi(\gamma_{N})
	-\psi(1-\gamma_{N})\right]\ =\ 1\ .
	\label{adsol}
\end{equation}

In order to interpret the $\Delta$-dependence of eq.~(\ref{adsol}),
it is useful to solve it as a power series in 
$\bar\alpha_{s}$ and then transform back to 
$x$-space.  The power series takes the form
\begin{equation}
	{\alpha_{s}\over2\pi}P_{gg}(x)\ =\ {\bar\alpha_{s}\over x}
	\sum_{m=0}C_{m}{\bar\alpha_{s}^{m}\left[\ln{1/x}-(m+1)
	\Delta\right]^{m}\over 	m!}\ ,
	\label{powser}
\end{equation}
where $C_{0}=1$, $C_{1}=C_{2}=0$, $C_{3}=2\zeta(3)$, etc.  As in the 
last section we see that this expansion is an asymptotic series, which 
can be best approximated by truncating at the largest value 
of $m$ for which $[\ln{1/x}-(m+1)\Delta]>0$.  The zeroth term in the 
expansion just corresponds to the standard double-logarithmic scaling 
of the DGLAP equation \cite{dglap}.  Thus, to see any effects of the resummation 
beyond double-logarithmic scaling, this expansion suggests that we 
must at the very least require
\begin{equation}
	x\lstsim e^{-4\Delta}\ .
	\label{req}
\end{equation}
This requirement is fairly strong, because the first nonzero correction 
occurs at $\alpha_{s}^{3}$.  For $\Delta=1$ and $\Delta=2$, this
gives $x\lstsim2\times10^{-2}$ and $x\lstsim3\times10^{-4}$, respectively, 
before one would expect to see some deviation from double-logarithmic 
scaling at small $x$.

\section{Modified BFKL Equation at NLL}
\label{sec:3}

As seen in the last two sections, the effects of the BFKL resummation 
can depend strongly on the rapidity-separation parameter $\Delta$, even 
though the 
dependence is formally NLL.  Furthermore, we argued in the introduction 
that a nonzero value of $\Delta$ seems appropriate, although it is not
obvious what is the best choice for this parameter.  Thus, we need to 
consider how to perform a NLL calculation, while retaining the dependence 
on $\Delta$.  In this section we show how to obtain the 
$\Delta$-dependence of the NLL kernel, and we explore its consequences.

The generalization of the $\Delta$-dependent BFKL equation takes
the form
\begin{eqnarray}
f( y_{a}- y_{b},q_{a\perp},q_{b\perp}) &=&  
{1\over2}\delta^{2}(q_{a\perp}-q_{b\perp})\Theta( y_{a}- y_{b}-\Delta)
\nonumber\\
&&
+\ \int_{ y_{b}+\Delta}^{ y_{a}-\Delta} 
d y\, K_{\Delta}\left[
f( y- y_{b},q_{a\perp},q_{b\perp})\right]\label{bfklnll} \ ,
\end{eqnarray}
where $K_{\Delta}$, which depends on $\Delta$, is an integral operator 
acting on the transverse momentum $q_{a\perp}$.
This operator can be expanded as a power series in $\bar\alpha_{s}$:
\begin{equation}
	K_{\Delta}\ =\ \bar\alpha_{s}K^{(1)}+
	\bar\alpha_{s}^{2}K^{(2)}+O(\bar\alpha_{s}^{3})\ ,
	\label{intop}
\end{equation}
where the first term $K^{(1)}$, which gives the LL equation, 
is $\Delta$-independent.  It can be read directly off of 
eq.~(\ref{bfkl}), yielding
\begin{equation}
K^{(1)}\left[f(q_{\perp})\right]\  =\ 
	 {1\over\pi}\int{d^{2}\ell_\perp
\over (\ell_{\perp}-q_{\perp})^{2}}\Biggl[
f(\ell_{\perp})
-{q_{\perp}^{2}\over \ell_{\perp}^{2}+(\ell_{\perp}-q_{\perp})^{2}}
f(q_{\perp})\Biggr] 
	\label{kone}
\end{equation}
for any function $f(q_{\perp})$.

The second term $K^{(2)}$ in eq.~(\ref{intop}) contains 
the NLL corrections to the BFKL kernel, and in the present 
formulation it will depend on the parameter $\Delta$.  The 
$\Delta$-dependence of this term can be found by the requirement that 
the total dependence on $\Delta$ of the high energy cross section 
must vanish up to NNLL terms.  That is, if we expand 
the hard cross section for scattering of partons $i$ and $j$,
\begin{equation}
	\hat\sigma_{NLL}^{ij}\ =\ \int d^{2}q_{a\perp}d^{2}q_{b\perp}\,
	V^{i}(q_{a\perp})\,
    f( y_{a}-y_{b},q_{a\perp},q_{b\perp})\,
	V^{j}(q_{b\perp})\ ,
	\label{xsec}
\end{equation}
in powers of $\alpha_{s}$, the coefficients of all terms of the form 
$\alpha_{s}^{b+1}(\alpha_{s}y)^{n}$ should be independent of $\Delta$, 
where the Born term is $O(\alpha_{s}^{b})$.  In this equation the 
quantities $V^{i}(q_{\perp})$ are the impact factors, which can also
be expanded in a power series in $\bar \alpha_{s}$:
\begin{equation}
	V^{i}(q_{\perp})\ =\ V^{i(0)}(q_{\perp})
	+\bar\alpha_{s}V^{i(1)}(q_{\perp})
	+O(\bar\alpha_{s}^{2})\ .
	\label{imps}
\end{equation}

The condition for $\Delta$-independence of the cross section
(\ref{xsec}) to NLL is now obtained by inserting the iterated solution 
to (\ref{bfklnll}) in (\ref{xsec}), using (\ref{intop}) and 
(\ref{imps}), and requiring that the coefficients of the terms of the 
form $\alpha_{s}^{b+1}(\alpha_{s}y)^{n}$ are independent of $\Delta$.
We find that the NLL kernel must be of the form
\begin{equation}
	K^{(2)}\left[f(q_{\perp})\right] 
	\ =\ 
	K^{(2)}\left[f(q_{\perp})\right] \bigg|_{\Delta=0}
	+\Delta\, 
	K^{(1)}\!\!\left[
	K^{(1)}\left[f(q_{\perp})\right] \right]\ ,
	\label{Ddep}
\end{equation}
where the first term is the $\Delta$-independent BFKL kernel 
given in Ref.~\cite{nll}, and all of the $\Delta$-dependence is in
 the second term.  Similarly, the NLL impact factors are of the form
\begin{equation}
	V^{i(1)}(q_{\perp})\ =\ V^{i(1)}(q_{\perp})\bigg|_{\Delta=0}
	+\Delta\, K^{(1)}\!\!\left[V^{i(0)}(q_{\perp})\right]\ .
	\label{Dimp}
\end{equation}
The virtual correction component of these results, 
(\ref{Ddep}) and (\ref{Dimp}), can also be obtained by 
considering the $\Delta$-dependent modification of the 
gluon-reggeization prescription, as discussed in the appendix.

At this stage we follow the lead of Ref.~\cite{nll} and
consider the action of the NLL kernel on the LL eigenfunctions, which 
have been modified so that the eigenvalue is symmetric in $\nu$.  
Specifically, we apply the operator (\ref{intop}) 
to the $n=0$ eigenfunction, which dominates at high-energy and gives the 
contribution to the total cross section.  We find
\begin{equation}
	K_{\Delta}\left[\alpha_{s}(q_{\perp}^{2})^{-1/2}\,
	(q_{\perp}^{2})^{-1/2+i\nu}\right]\ =\ 
	\omega(\nu)\,\alpha_{s}(q_{\perp}^{2})^{-1/2}\,
	(q_{\perp}^{2})^{-1/2+i\nu}\
	\label{eigii}
\end{equation}
with
\begin{equation}
	\omega(\nu)\ =\ \omega^{(0)}(\nu)\left(1
	-{\bar\alpha_{s}\over4}\bar c(1/2+i\nu)\right)
	+\Delta\,\left(\omega^{(0)}(\nu)\right)^{2}\ .
	\label{eigiv}
\end{equation}
In this equation $\omega^{(0)}(\nu)$ is the LL eigenvalue 
with running coupling:
\begin{equation}
	\omega^{(0)}(\nu)\ =\ 
	2\bar\alpha_{s}(q_{\perp})\left[\psi(1)-{\rm Re}\,
	\psi\left({1\over2}+i\nu\right)\right]	\ ,
	\label{eigiii}
\end{equation}
while the function $\bar c(\gamma)$ contains the $\Delta$-independent 
NLL corrections to the eigenfunction.  The exact expression for 
$\bar c(\gamma)$ can be obtained from the function $c(\gamma)$ in 
Ref.~\cite{nll} by removing the terms antisymmetric in $\nu$.  
The last term in (\ref{eigiv}) is the modification of the NLL 
eigenfunction due to the rapidity-separation constraint.

We now consider the solution of the modified BFKL equation at NLL.
For simplicity we will ignore the effects of the running coupling%
\footnote{Running coupling in NLL BFKL has been considered in 
Refs.~\cite{mueller,levin,bartels}.  It produces important effects 
such as non-Regge terms in high-energy cross sections.  However, the 
running of the coupling appears to be somewhat independent of the 
large scale-invariant corrections which are the main concern here.}.
Then, following the same procedure as in section 2,
 one obtains the BFKL Green's function solution (\ref{sold}) 
 as an integral over $\nu$  with the coefficient of the rapidity in 
the exponent given by the implicit solution to
\begin{equation}
	\tilde\omega(\nu)\ =\ \omega(\nu)e^{-\tilde \omega(\nu)\Delta}\ .
	\label{tilop}
\end{equation}
The saddle-point approximation to the integral (\ref{sold}) is 
determined in terms of the expansion of $\tilde\omega(\nu)$ 
around $\nu=0$:
\begin{equation}
	\tilde\omega(\nu)\ =\ \tilde A-\tilde B\nu^{2}+\cdots\ ,
	\label{nlltaylor}
\end{equation}
where the coefficients $\tilde A$ and $\tilde B$ are related to the 
equivalent coefficients $A$ and $B$ in the expansion of $\omega(\nu)$ by
\begin{eqnarray}
	\tilde A & = & Ae^{-\tilde A\Delta}\nonumber  \\
	\tilde B & = & {B\,(\tilde A/A)\over1+\tilde A\Delta}\ .
	\label{tilabi}
\end{eqnarray}
The values of $A$ and $B$ are
\begin{equation}
	A\ =\ 2.77\bar\alpha_{s}+(-18.34+7.69\Delta)\bar\alpha_{s}^{2}
	\label{tilopi}
\end{equation}
and
\begin{equation}
	B\ =\ 16.83\bar\alpha_{s}+(-321.49+93.32\Delta)\bar\alpha_{s}^{2}
	\label{tilopii}
\end{equation}
for three active flavors.

The coefficient $\tilde A$ is related to the BFKL Pomeron intercept by
$\tilde A=\omega_{P}-1$.  In figure 2 we plot both the LL and the NLL
predictions for $\tilde A$ as a function of $\Delta$ for 
$\alpha_{s}=0.15$.
We note that, although the NLL corrections to $\tilde A$ are large for
$\Delta=0$, they are not large for $\Delta\gtrsim 2$ and they vanish 
for $\Delta=2.4$.  Furthermore, the dependence of the NLL solution 
on $\Delta$ is very weak for $\Delta\gtrsim 2$.  

\begin{figure} 
\vskip1.0cm
\epsfysize=10.0cm
\centerline{\epsffile{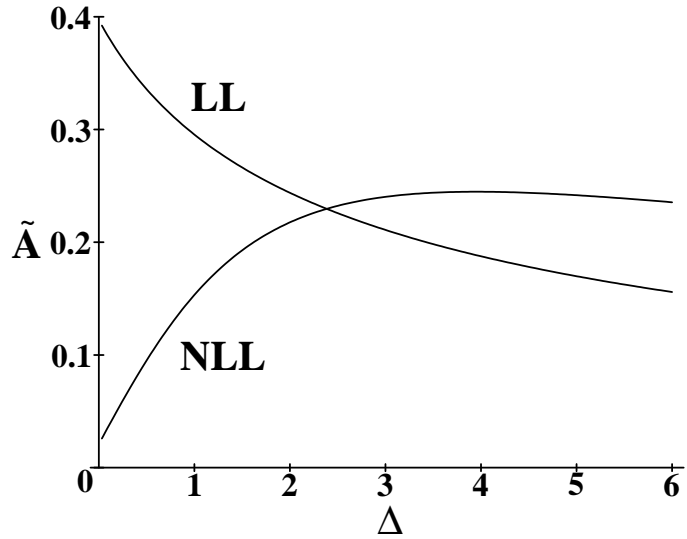}}
\vskip-1.5cm
\vskip6pt
\caption{BFKL Pomeron intercept $\tilde A=\omega_{P}-1$ at LL and NLL 
    as a function of $\Delta$ for $\alpha_{s}=0.15$.}
\label{fig2}
\vskip-1.0cm
\end{figure}

\begin{figure} 
\vskip1.0cm
\epsfysize=10.0cm
\centerline{\epsffile{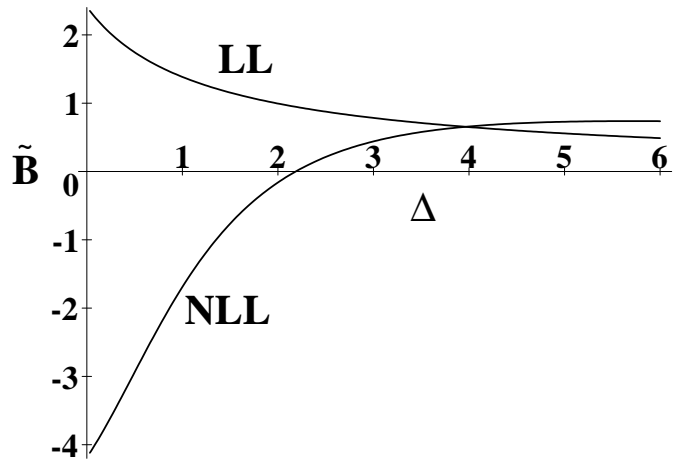}}
\vskip-1.5cm
\vskip6pt
\caption{The coefficient $\tilde B$ of eq.~(\ref{nlltaylor}) at LL and NLL 
    as a function of $\Delta$ for $\alpha_{s}=0.15$.}
\label{fig3}
\vskip-1.0cm
\end{figure}

In figure 3 we plot the coefficient $\tilde B$ at LL and NLL as a 
function of $\Delta$ for $\alpha_{s}=0.15$.  At $\Delta=0$ it is 
negative indicating that the standard BFKL eigenvalue has a 
minimum at $\nu=0$ rather than a maximum \cite{ross}.  It has been 
suggested that this 
leads to disastrous consequences such as oscillations in the cross 
section \cite{levin}. At the very least it shows that the 
interpretation of $\tilde A$ as the Pomeron intercept $\omega_{P}-1$
is invalid at $\Delta=0$.
However, for $\Delta\gtrsim2.2$ the coefficient $\tilde B$ becomes 
positive again, so the modified solution 
does have a maximum at $\nu=0$, and we can again identify $\tilde A$ 
with the BFKL Pomeron intercept.  This suggests that a 
value of the rapidity-separation parameter of $\Delta\gtrsim 2.2$ is 
more natural than the standard choice of $\Delta=0$.  On this basis, 
we estimate the value of the NLL BFKL Pomeron intercept to be 
between 0.22 and 0.25 for $\alpha_{s}=0.15$.

\section{Conclusions}
\label{sec:conclusions}

In this paper we have considered the modification of the BFKL 
resummation by requiring that the rapidities of 
successively emitted gluons must satisfy the constraint 
$y_{i+1}-y_i>\Delta$.  The inclusion of an arbitrary 
``renormalization'' constant, such 
as $\Delta$, in the BFKL resummation is a natural thing to do, because 
in such a resummation of large logarithms one can always redefine the 
energy scale in the logarithm.  The particular 
implementation of this arbitrary constant (the 
``renormalization scheme'') as a constraint on the 
rapidity separations is nice because it has an obvious physical 
interpretation.  Using this interpretation, we have argued 
and we have seen through 
specific calculations that the choice of $\Delta=0$ is not the best 
choice for performing the resummation.  This is analogous 
to performing a fixed-order calculation using an inappropriate choice 
of the ultraviolet 
renormalization scale $\mu$.  The next-to-leading corrections 
are large, not because the perturbative calculation is 
inherently bad, but because one has simply made a poor choice of scale.

We have shown how to consistently include the rapidity-separation 
constraint, both at LL and at NLL. At LL we find that the prediction 
for the BFKL Pomeron intercept decreases monotonically as $\Delta$ 
increases, while at NLL it increases rapidly from $\Delta=0$ and then
becomes quite insensitive to $\Delta$ for $\Delta\gtrsim2$ and 
$\alpha_{s}=0.15$.  Furthermore, the NLL corrections are relatively 
small compared to the LL prediction 
for $\Delta\gtrsim2$.  We have also seen that the eigenvalue has a maximum 
at $\nu=0$ as long as $\Delta\gtrsim2.2$, so that the bad behavior 
seen in the saddle-point approximation is no longer a problem.
It is interesting to note 
that our results for large $\Delta$ are in reasonable agreement with the 
results of Ref.~\cite{salam}, which addresses the question of the large 
NLL corrections by resumming double logarithms in transverse momenta.
Presumably, the same sort of correlation effects are being included by 
the two very different approaches.

In conclusion, we believe that for a reasonable range of the 
rapidity-separation 
parameter $\Delta$ the modified BFKL formalism is a theoretically 
consistent and stable resummation of the perturbation series in 
$\alpha_{s}$ at high-enough energies.  However, the phenomenological 
usefulness of this resummation is still an open question.  If we 
assume that $\Delta>2$ is required to consistently use the modified-BFKL 
resummation, then one should not expect to see significant deviations from 
double-logarithmic scaling at small-$x$ in the gluon-gluon splitting 
functions at least until $x\lstsim3\times10^{-4}$.  
Similarly, equation (\ref{expom}) suggests that in large-rapidity dijet 
production one needs rapidity intervals of $y\gtrsim4$ before the 
BFKL formalism would begin to be applicable\footnote{
In fact, for the measurement of the hadron-collider 
energy dependence of the two-jet cross section at fixed Feynman-$x$'s, 
as suggested by Mueller and Navelet \cite{mn}, 
one would need $y\gtrsim6$, because the first non-trivial term in the 
perturbative expansion in $\alpha_{s}y$ vanishes.}.
Certainly, more detailed phenomenological study, using the insights 
gained from the NLL corrections, is needed in order to assess the 
importance of BFKL to experiment.

\vspace{.5cm}

{\sl Acknowledgments}
We would like to thank Vittorio Del Duca and Wu-Ki Tung for useful 
comments on this manuscript.

\vspace{.5cm}

\appendix
\section{Modification of the Reggeized Gluons}
\label{sec:appa}

An important requirement of our analysis is that the modification of 
the real and virtual gluon rapidity phase space is handled 
identically, so that the cancellation of the soft singularities 
remains intact.  However, it is still possible to include the virtual 
corrections using a reggeized gluon, although with a slightly 
more complicated, $\Delta$-dependent form.  For the modified BFKL 
equation (\ref{bfklnll}) we find that the appropriate prescription 
for reggeizing a gluon of off-shellness $t_{i}$
is to replace the gluon propagator by
\begin{equation}
	{1\over t_{i}}\rightarrow {1\over 
	t_{i}}\,{e^{-\tilde\lambda(t_{i})\Delta/2}\over
	\sqrt{1+\tilde\lambda(t_{i})\Delta}}\,
	\left({s_{i+1,i}\over|k_{i+1\perp}||k_{i\perp}|}
	\right)^{\tilde\lambda(t_{i})/2}\ =\ 
	{1\over t_{i}}\,
	{e^{\tilde\lambda(t_{i})(y_{i+1,i}-\Delta)/2}
	\over\sqrt{1+\tilde\lambda(t_{i})\Delta}}\,
	\label{regglue}
\end{equation}
where $\tilde\lambda(t)$ is related to the usual BFKL Regge intercept 
$\alpha(t)\equiv\lambda(t)/2$ by solving the equation
\begin{equation}
		\tilde\lambda(t)\ = \ \lambda(t)e^{-\tilde\lambda(t)\Delta}\ .
	\label{modreg}
\end{equation}
Note that the square root in the equation (\ref{regglue}) occurs 
because the rapidity phase space modification is defined at the 
squared-amplitude level, whereas the reggeization is defined at the 
amplitude level.  

The quantity $\lambda(t)$ is just the purely virtual 
contribution to the kernel $K_{\Delta}$.
It can be expanded as a power series in 
$\bar\alpha_{s}$:  
\begin{equation}
	\lambda(t)\ =\ \bar\alpha_{s}\lambda^{(1)}(t)+\bar\alpha_{s}^{2}
	\lambda^{(2)}(t)+O(\bar\alpha_{s}^{3})\ ,
	\label{alex}
\end{equation}
where the $\Delta$-dependence begins with the $\lambda^{(2)}(t)$ term.
At LL the kernel can be separated into a real and a virtual 
contribution:
\begin{equation}
	K^{(1)}\left[f(q_{\perp})\right]\ =\ K^{(1)}_{r}\left[f(q_{\perp})\right]
	+K^{(1)}_{v}\left[f(q_{\perp})\right]\ ,
	\label{llrv}
\end{equation}
where 
\begin{equation}
	K^{(1)}_{v}\left[f(q_{\perp})\right]\ =\ 
	\lambda^{(1)}(-q_{\perp}^{2})\,f(q_{\perp})\ ,
	\label{veig}
\end{equation}
and $t=q^{2}\simeq-q_{\perp}^{2}$.  Thus, $\lambda^{(1)}$ can be 
obtained directly from (\ref{kone}):
\begin{eqnarray}
\lambda^{(1)}(-q_{\perp}^{2}) &=&-{1\over\pi(2\pi)^{-2\epsilon}}
\int d^{2-2\epsilon}
\ell_\perp\, {q_{\perp}^{2}\over (\ell_{\perp}-q_{\perp})^{2}
\left(\ell_{\perp}^{2}+(\ell_{\perp}-q_{\perp})^{2}\right)}\nonumber\\
&=&-{1\over(2\pi)^{1-2\epsilon}}\int d^{2-2\epsilon}
\ell_\perp\, {q_{\perp}^{2}\over \ell_{\perp}^{2}
(\ell_{\perp}-q_{\perp})^{2}}
	\label{aone}\\
&=&{1\over\epsilon} 
\left(\mu^2\over q_{\perp}^{2}\right)^{\epsilon} {\Gamma(1+\epsilon)\,
\Gamma^2(1-\epsilon)\over (4\pi)^{-2\epsilon}\,\Gamma(1-2\epsilon)}\ .
\nonumber
\end{eqnarray}
We also obtain
\begin{equation}
	K^{(1)}_{r}\left[f(q_{\perp})\right]\ =\ 
	{1\over\pi(2\pi)^{-2\epsilon}}\int d^{2-2\epsilon}
\ell_\perp\, {f(\ell_{\perp})\over (\ell_{\perp}-q_{\perp})^{2}}\ .
	\label{reig}
\end{equation}
For definiteness we have used dimensional
regularization with $d=4-2\epsilon$ dimensions to render the integrals 
finite.

At NLL the kernel can be separated into the real and virtual 
corrections to the Lipatov vertex $K^{(1)}_{r}$ plus the two-loop 
virtual contribution:
\begin{equation}
	K^{(2)}\left[f(q_{\perp})\right]\ =\ K^{(2)}_{rr}\left[f(q_{\perp})\right]
	+K^{(2)}_{rv}\left[f(q_{\perp})\right]
	+K^{(2)}_{vv}\left[f(q_{\perp})\right]\ ,
	\label{llrvv}
\end{equation}
where 
\begin{equation}
	K^{(2)}_{vv}\left[f(q_{\perp})\right]\ =\ 
	\lambda^{(2)}(-q_{\perp}^{2})\,f(q_{\perp})\ .
	\label{veigii}
\end{equation}
Similarly, the NLL impact factors can be separated into real and 
virtual corrections:
\begin{equation}
	V^{i(1)}(q_{\perp})\ =\ V^{i(1)}_{r}(q_{\perp})
	+V^{i(1)}_{v}(q_{\perp})\ .
	\label{imprvv}
\end{equation}

It is now straightforward to verify the $\Delta$-dependence of the NLL
kernel and impact factors given in equations (\ref{Ddep}) and (\ref{Dimp}),
at least for the virtual components $K^{(2)}_{vv}$,  $K^{(2)}_{rv}$, 
and $V^{i(1)}_{v}$.  This is most easily seen by reorganizing the 
reggeized-gluon propagator (\ref{regglue}) using
\begin{eqnarray}
	{e^{\tilde\lambda(t)(y-\Delta)}
	\over{1+\tilde\lambda(t)\Delta}}
	&=& \left[1-2\bar\alpha_{s}
	\Delta\lambda^{(1)}(t)\right]
	\exp\left\{\left[\bar\alpha_{s}\lambda^{(1)}(t)+\bar\alpha_{s}^{2}
	[\lambda^{(2)}(t)-\Delta(\lambda^{(1)}(t))^{2}]\right]y\right\}\nonumber\\
	&&+\ {\cal O}\left(\bar\alpha_{s}^{2}[\bar\alpha_{s}y]^{n}\right)\ ,
	\label{regglueii}
\end{eqnarray}
where the remaining terms are effectively NNLL.  Demanding 
$\Delta$-independence up to NNLL, one immediately obtains from the 
term in the exponent
\begin{equation}
	\lambda^{(2)}(t)
	\ =\ 
	\lambda^{(2)}(t) \bigg|_{\Delta=0}
	+\Delta\, 
	\left(\lambda^{(1)}(t)\right)^{2}\ .
	\label{Ddepvv}
\end{equation}
Using the modified form of the reggeized gluon in a $2\rightarrow n$ 
partons high-energy amplitude and expanding to one-loop order, 
as done in Refs.~\cite{dds,ddsii}, one also directly obtains
\begin{equation}
	V^{i(1)}_{v}(q_{\perp})\ =\ V^{i(1)}_{v}(q_{\perp})\bigg|_{\Delta=0}
	+\Delta\, \lambda^{(1)}(-q_{\perp})\,V^{i(0)}(q_{\perp})\ ,
	\label{Dimpv}
\end{equation}
and
\begin{eqnarray}
	K^{(2)}_{rv}\left[f(q_{\perp})\right] 
	&=& 
	K^{(2)}_{rv}\left[f(q_{\perp})\right] \bigg|_{\Delta=0}\nonumber\\
	&&
	+\,\Delta \left(
	K^{(1)}_{r}\left[\lambda^{(1)}(-q_{\perp}^{2})f(q_{\perp})\right]
	+\lambda^{(1)}(-q_{\perp}^{2})\,K^{(1)}_{r}\left[f(q_{\perp})\right]
	\right)\ .
	\label{Ddeprv}
\end{eqnarray}
The equations (\ref{Ddepvv}), (\ref{Dimpv}), and  (\ref{Ddeprv})
are just the virtual-correction components of equations (\ref{Ddep}) 
and (\ref{Dimp}).


\begin{thebibliography}{99}

\bibitem{FKL} E.A.~Kuraev, L.N.~Lipatov and V.S.~Fadin, 
{\em Zh.~Eksp.~Teor.~Fiz.}
{\bf 71}, 840 (1976) [{\em Sov.~Phys.~JETP} {\bf 44}, 443 (1976)].

\bibitem{fkl2} E.A.~Kuraev, L.N.~Lipatov and V.S.~Fadin, {\em 
Zh.~Eksp.~Teor.~Fiz.} {\bf 72}, 377 (1977) [{\em Sov.~Phys.~JETP} {\bf 45}, 
199 (1977)].

\bibitem{bal} Ya.Ya.~Balitsky and L.N.~Lipatov, {\em Yad.~Fiz.} {\bf 28} 
1597 (1978) [{\em Sov.~J.~Nucl.~Phys.} {\bf 28}, 822 (1978)].

\bibitem{nll} V.S.~Fadin and L.N.~Lipatov, {\em Phys.~Lett.} {\bf B429}, 
127 (1998). 

\bibitem{fadin} V.S.~Fadin, preprint {\tt hep-ph/9807528}.

\bibitem{real} V.S.~Fadin and L.N.~Lipatov, {\em Yad.~Fiz.} {\bf 50}, 
1141 (1989)
[{\em Sov.~J.~Nucl.~Phys.} {\bf 50}, 712 (1989)]; {\em Nucl.~Phys.} 
{\bf B477}, 767 (1996).

\bibitem{fl} V.S.~Fadin and L.N.~Lipatov, {\em Nucl.~Phys.} {\bf
B406}, 259 (1993).

\bibitem{virtual} V.S.~Fadin, R.~Fiore and 
A.~Quartarolo, {\em Phys.~Rev.} {\bf D50} 5893 (1994).

\bibitem{linear} V.S.~Fadin, R.~Fiore and M.I.~Kotsky, {\em 
Phys.~Lett.} 
{\bf B389}, 737 (1996).

\bibitem{ffk} V.S.~Fadin, R.~Fiore and M.I.~Kotsky, {\em Phys.~Lett.} 
{\bf B359}, 181 (1995); {\bf B387} 593 (1996);\\ V.S.~Fadin, 
R.~Fiore and A.~Quartarolo, {\em Phys.~Rev.} {\bf D53}, 2729 (1996);\\
J.~Bl\"umlein, V.~Ravindran, W.L.~van Neerven, {\em Phys.~Rev.} {\bf D58},
09152 (1998).

\bibitem{vdd} V.~Del~Duca, {\em Phys.~Rev.} {\bf D54}, 989, 4474 (1996).

\bibitem{dds} V.~Del~Duca and C.R.~Schmidt, {\em Phys.~Rev.} {\bf D57}, 4069
(1998).

\bibitem{ff}V.S.~Fadin and R.~Fiore, {\em Phys.~Lett.} {\bf B294}, 286 (1992);\\
V.S.~Fadin, R.~Fiore and A.~Quartarolo, {\em Phys.~Rev.}
{\bf D50}, 2265 (1994).

\bibitem{ddsii} V.~Del~Duca and C.R.~Schmidt, preprint EDINBURGH 
98/21, MSUHEP-80928, {\tt hep-ph/9810215}.

\bibitem{bds} Z.~Bern, V.~Del~Duca and C.R.~Schmidt, preprint 
EDINBURGH 98/20, MSUHEP-81016, UCLA/98/TEP/29, {\tt hep-ph/9810409}.

\bibitem{cc} G.~Camici and M.~Ciafaloni, {\em Phys.~Lett.} {\bf B412},
396 (1997); Erratum-ibid. {\bf B417} 390 (1998); {\em Phys.~Lett.} {\bf 
B430}, 349 (1998).

\bibitem{ross} D.A.~Ross, {\em Phys.~Lett.} {\bf B431}, 161 (1998).

\bibitem{levin} E.~Levin, preprint TAUP 2501-98, {\tt hep-ph/9806228}.

\bibitem{ball} R.D.~Ball and S.~Forte, preprint {\tt 
hep-ph/9805315};\\
J.~Bl\"umlein {\it et al,} {\em Phys.~Rev.} {\bf D58}, 014020
(1998); preprint {\tt hep-ph/9806368}.

\bibitem{bo} B.~Andersson, G.~Gustafson, J.~Samuelsson, {\em Nucl.~Phys.} 
{\bf B467}, 443 (1996).

\bibitem{liptalk} L.N.~Lipatov, talk presented
at the $4^{\rm th}$ Workshop on Small-$x$ and Diffractive Physics,
Fermi National Accelerator Laboratory, Sept. 17-20, 1998.

\bibitem{carl} C.R.~Schmidt, {\em Phys.~Rev.~Lett.} {\bf 78}, 4531 (1997).

\bibitem{os} L.H.~Orr and W.J.~Stirling, {\em Phys.~Rev.} {\bf D56}, 
5875 (1997).

\bibitem{cathaut} S.~Catani and F.~Hautmann, {\em Nucl.~Phys.} 
{\bf B427}, 475 (1994).

\bibitem{dglap} Y.L.~Dokshitzer, {\em Sov.~Phys.~JETP} {\bf 73}, 1216 
(1977);\\
V.N.~Gribov and L.N.~Lipatov, {\em Sov.~J.~Nucl.~Phys.} {\bf 15}, 78 
(1972);\\
G.~Altarelli and G.~Parisi, {\em Nucl.~Phys.} {\bf B126}, 298 (1977).

\bibitem{mueller} Y.V.~Kovchegov and A.H.~Mueller,
{\em Phys.~Lett.} {\bf B439}, 428 (1998).

\bibitem{bartels} N.~Armesto, J.~Bartels, M.A.~Braun, 
{\em Phys.~Lett.} {\bf B442}, 459 (1998).

\bibitem{salam} G.P.~Salam, {\em J.High Energy Phys.} {\bf 9807}, 019 
(1998). 

\bibitem{mn} A.H.~Mueller and H.~Navelet, {\em Nucl.~Phys.} 
{\bf B282}, 727 (1987).

\end{thebibliography}
\end{document}